\documentclass[twocolumn,showpacs,preprintnumbers,amsmath,amssymb,superscriptaddress]{revtex4}
\usepackage{graphicx}
\usepackage{dcolumn}
\usepackage{bm}
\usepackage[latin1]{inputenc}
\usepackage{color}

\begin{document}

\preprint{APS/123-QED}

\title{Taylor-like vortices in shear-banding flow of giant micelles}

\author{M.A. Fardin}
\affiliation{Laboratoire Mati\`ere et Syst\`emes Complexes, Universit\'e Paris
7--CNRS UMR 7057,\\10 rue Alice Domon et L\'eonie Duquet, 75205 Paris C\'edex 13, France.}%
\author{B. Lasne}
\affiliation{Laboratoire Mati\`ere et Syst\`emes Complexes, Universit\'e Paris
7--CNRS UMR 7057,\\10 rue Alice Domon et L\'eonie Duquet, 75205 Paris C\'edex 13, France.}%
\author{O. Cardoso}
\affiliation{Laboratoire Mati\`ere et Syst\`emes Complexes, Universit\'e Paris
7--CNRS UMR 7057,\\10 rue Alice Domon et L\'eonie Duquet, 75205 Paris C\'edex 13, France.}%
\author{G. Grégoire}
\affiliation{Laboratoire Mati\`ere et Syst\`emes Complexes, Universit\'e Paris
7--CNRS UMR 7057,\\10 rue Alice Domon et L\'eonie Duquet, 75205 Paris C\'edex 13, France.}%
\author{M. Argentina}%
\affiliation{ Laboratoire J.A. Dieudonn\'e, Universit\'e de Nice-Sophia Antipolis,\\ Parc Valrose, F-06108  Nice C\'edex 02, France.} 
\author{J.P. Decruppe}
\affiliation{Laboratoire de Physique des Milieux Denses, Universit\'e Paul Verlaine,\\ 1 Bd Arago, 57078 Metz C\'edex 02, France.}%
\author{S. Lerouge}
\altaffiliation[Corresponding author ]{}
\email{sandra.lerouge@univ-paris-diderot.fr}
\affiliation{Laboratoire Mati\`ere et Syst\`emes Complexes, Universit\'e Paris
7--CNRS UMR 7057,\\10 rue Alice Domon et L\'eonie Duquet, 75205 Paris C\'edex 13, France.}%

\date{\today}

\begin{abstract}

Using flow visualizations in Couette geometry, we demonstrate the existence of Taylor-like vortices in the shear-banding flow of a giant micelles system. We show that vortices stacked along the vorticity direction develop  concomitantly with interfacial undulations. These cellular structures are mainly localized in the induced band and their dynamics is fully correlated to that of the interface. As the control parameter increases, we observe a transition from a steady vortex flow to a state where pairs of vortices are continuoulsy created and destroyed. Normal stress effects are discussed as potential mechanisms driving the three-dimensional flow. 
 
\end{abstract}

\maketitle

\begin{figure}[t]
\begin{center}
\includegraphics{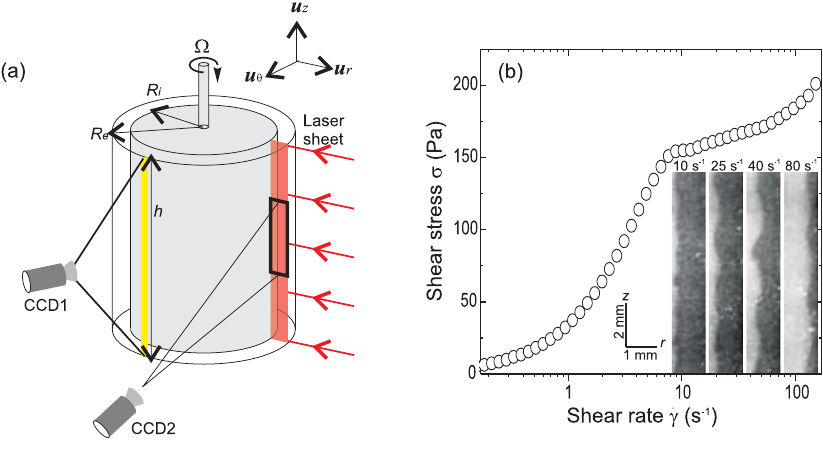}
\caption{(a) Experimental setup for observations in the ($\mathbf{u}_{\theta},\mathbf{u}_{z}$) and ($\mathbf{u}_{r},\mathbf{u}_{z}$) planes. (b) Steady-state flow curve and corresponding banding structure in the ($\mathbf{u}_{r},\mathbf{u}_{z}$) plane for different shear rates along the stress plateau. The left and right sides of each picture corresponds respectively to the inner and outer cylinders.
\label{fig1}}
\end{center}
\end{figure}

\begin{figure}[b]
\includegraphics{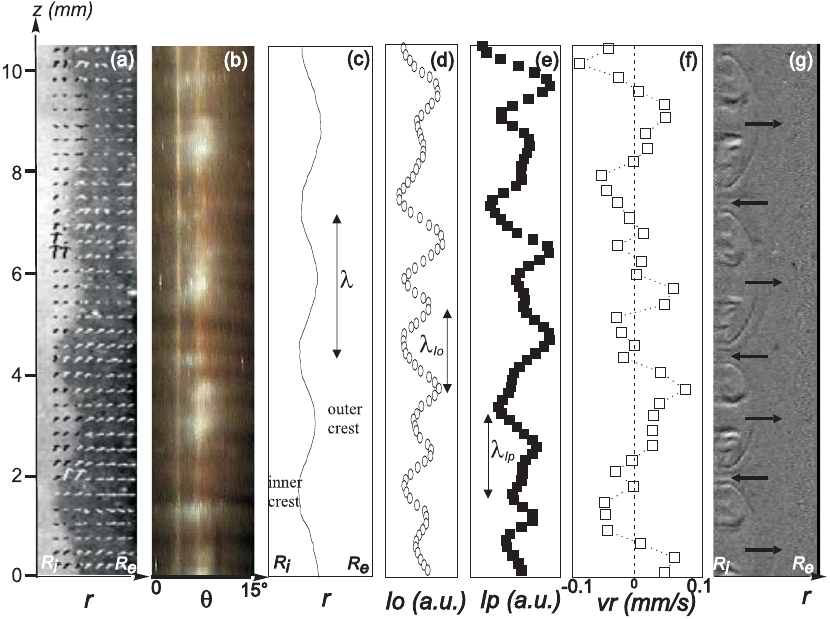}
\caption{(a) Stationary banding structure at $\dot\gamma$=40 s$^{-1}$ and corresponding velocity field computed from the PIV algorithm in the ($r,z$) plane. (b) Side view of the Couette cell under ambient illumination.  (c) Interface profile $I(z)$ extracted from (a). (d) Intensity distribution $I_{o}(z)$ along the vertical axis inferred from (b). (e) Mean orientation $I_{p}(z)$ of velocity vectors with respect to the vertical axis. (f) Radial velocity profile $v_{r}$ at a position $r=R_{i}+0.5e$ in the gap highlighting the inflow and outflow. (g) Direct observation in the plane ($r,z$) of vortices stacked along the vorticity direction. In this last case, the sample is free of tracers and illuminated with white light. \label{fig2}}
\end{figure}

In contrast to Newtonian fluids, complex fluids are structured at the mesoscopic scale, typically from 10 nm to 100 $\mu$m, corresponding, for example, to  the mesh size in entangled polymers solutions or to the bubble size in foams. Such a length scale provides internal degrees of freedom with characteristic times in the order of 10$^{-3}$ to 100 s (position, orientation, composition, ...) that can easily be excited by the flow, often resulting in a new organization of the structure of the fluid and of the flow itself. The relation between stress and shear rate $\sigma=f(\dot\gamma)$ becomes strongly nonlinear and the flow evolves towards a spatially heterogeneous shear-banded state where regions of differing viscosities coexist. The shear-banding phenomenon is ubiquitous in complex fluids, from surfactants solutions \cite{Berret,Salmon}, to telechelic polymers \cite{Manneville2007,Sprakel2008}, emulsions \cite{Coussot,Becu2006}, granular materials \cite{Losert} and foams \cite{Gilbreth}. Among these systems, surfactant giant micelles have generated an abundant literature \cite{Berret} and are often considered as model systems for the study of shear-banded flows. The  $\sigma=f(\dot\gamma)$ curve is made of two stable shear-thinning branches, characteristic of two distinct states of the system, separated by a stress plateau ($\sigma=\sigma_{p}$) that extends from $\dot\gamma_{l}$ to $\dot\gamma_{h}$. In the plateau regime, the flow is unstable and evolves towards a banded state where the  viscous and fluid phases coexist at constant  stress $\sigma_{p}$ \cite{Cates96}. This picture is supported, among others, by investigations focused on the determination of the flow field \cite{Salmon2003,Hu2005,Manneville2008,Callaghan2008}, revealing a discontinuity of the velocity profile in the flow plane. These 1D measurements also allowed to establish the existence of large spatio-temporal fluctuations of the local flow field \cite{Becu2007,Lopez2004}, suggesting the presence of three-dimensional flow \cite{Becu2007}. Recently, using the scattering properties of the induced band, we showed that the interface between bands is not flat as usually assumed but undulates along the vorticity direction \cite{Lerouge2006,Lerouge2008}. Such a scenario has been  qualitatively reproduced by a numerical perturbation analysis of the modified Johnson-Segalman equation  \cite{Fielding2007}. In addition, the model predicts the formation of velocity rolls stacked along the vorticity direction.\\
In this letter, we report the first direct observation of Taylor-like vortices in the shear-banding flow of a semi-dilute micellar solution. We follow simultaneously the interface and flow dynamics and show that, as the interface between bands becomes unstable, a secondary axisymmetric vortex flow develops. The Taylor-like vortices are stacked along the vorticity axis and are mainly located in the high shear band, their size depending on the applied shear rate. The radial velocities are found to be two orders of magnitude lower than the azimuthal base flow. Using flow visualizations, we show that the interface and flow dynamics are strongly correlated~:~as the control parameter increases, we observe a transition between a steady vortex flow towards a state dominated by continuous creations and  annihilations of cellular structures.\\
In our experiments, the working fluid is a 11\%~wt. semi-dilute aqueous mixture of surfactant (cetyltrimethylammonium bromide) and salt (sodium nitrate) at 28\r{}C. It is contained in a transparent Perspex cylindrical Couette device (gap $e$=1.13 mm, inner radius $R_i$=13.33 mm, height $h$=40 mm) designed for imaging of the flow gradient-vorticity plane and side views of the cell (Fig.~\ref{fig1}.a). The device is attached to the shaft of a commercial rheometer (Physica MCR500), with the inner cylinder rotating. The experiments are performed with the shear rate as the control parameter and the response of the system is recorded from two distinct channels, in a simultaneous way. 
The first one is dedicated to flow vizualizations in the flow-vorticity ($\mathbf{u}_{\theta},\mathbf{u}_{z}$) plane using seeding anisotropic reflective particles (anisotropic mica platelets from Merck at a volume fraction of 6.10$^{-5}$), as it was done for the study of the dynamic regimes of inertial Taylor-Couette flows \cite{Andereck}. The fluid is illuminated by ambient light and the intensity $I_{o}(z)$ reflected in the velocity gradient direction is collected on a digital camera (CCD1). 
The second channel is dedicated to the observation of a cross-section (($\mathbf{u}_{r},\mathbf{u}_{z}$) plane), using either a He-Ne laser sheet ($\lambda$=633 nm, thickness 1.5 mm) propagating along the velocity gradient axis or white light. A digital camera (CCD2) records the scattered intensity at 90$^{\circ}$. This channel provides information on the interface dynamics and allows the computation of the flow field using particle image velocimetry (PIV). 
Figure \ref{fig1}.b shows the steady-state shear stress as a function of the shear rate for the sample sedded with mica flakes. This flow curve, measured in strain-controlled mode, exhibits a stress plateau with a significant positive slope at  $\sigma_{p}=150\pm 1$ Pa, extending from $\dot\gamma_{l}=7\pm 0.5s^{-1}$ to $\dot\gamma_{h}=95\pm 5s^{-1}$. Along the stress plateau, the system is  organized into two bands separated by an interface that undulates along the vorticity direction with a well-defined wavelength  \cite{Lerouge2006,Lerouge2008}. The proportion of the induced band and the wavelength of the undulations are increasing functions of the applied shear rate (inset in Fig.~\ref{fig1}.b).
The flow structure associated with the interfacial undulations is illustrated in Fig.~\ref{fig2} for $\dot\gamma$=40 s$^{-1}$. On the snapshot of a cross-section taken when steady-state is achieved, we overlay the corresponding instantaneous flow field computed using PIV (see Fig.~\ref{fig2}.a). If, at first sight, no particular pattern can be extracted, axial and radial velocity components are detected, suggesting that the flow is not purely orthoradial. The existence of a secondary flow is confirmed by side views of the Couette cell under ambient illumination that reveals a succession of axisymmetric dark and bright stripes stacked along the vorticity axis (Fig.~\ref{fig2}.b). 
Such a picture is reminiscent of the primary pattern of the classical Taylor instability \cite{Andereck} and points out   that the base flow is replaced by a vortex flow. In fact, the variations of the  reflected intensity results from differences in the orientation of the seeding flow-aligned platelets~:~dark stripes are associated with regions where the flakes do not reflect the light  along the observation direction, indicating radial flow, while bright stripes correspond to flow perpendicular to the observation direction \cite{Savas}. In other words, the vortices axis are centred on bright stripes, the radial flow coming from the adjacent dark stripes. The comparison between the reflected light distribution $I_o(z)$ and the interface profile noted $I(z)$ (Fig.~\ref{fig2}.d and c) shows that the flow is oriented radially at the crests level of the interface profile. This means that, the interfacial wavelength $\lambda$ is twice the wavelength $\lambda_{I_o}$ characterizing the axial structure of the flow. To differentiate the inflow and outflow and quantify the corresponding velocities, we use the flow field infered from PIV. As mentioned above, due to experimental limitations, it is not possible to distinguish cellular structures from our measured flow field. Nonetheless, to test its validity, we compute the  following quantity that provides a measurement of the mean local orientation of velocity vectors with respect to the vorticity direction~:~$I_{p}(z)=\sum_{i}{\left(\left|v_z(z,r_i)\right|/\sqrt{v_z^2(z,r_i) + v_r^2(z,r_i)}\right)}$. Taking into account the relation between the intensity profile $I_o(z)$ and the radial velocity component \cite{Mutabazi}, we expect the mean orientation profile $I_p(z)$ to be similar to $I_o(z)$. Fig.~\ref{fig2}.(d-e) show that the axial structure $I_{p}(z)$ is consistent with the intensity profile $I_{o}(z)$ gathered from side views of the Couette cell, indicating that, even if our computed instantaneous flow field is noisy, some relevant information can be extracted from the latter. Fig.~\ref{fig2}.f displays the radial component $v_{r}(z)$ at a given position close to the interface between bands. First, we find that the outflow and inflow  are aligned respectively with the outer and inner crests of the interface profile. Second, radial velocities are in the order of 0.1 mm.s$^{-1}$, namely at least two orders of magnitude lower than the mean base azimuthal velocity. The weakness of the secondary flow makes the determination of the tracers displacement difficult and explains the absence of pattern in our mapping of the local velocities in the cross-section.\\ 
From those experiments we can conclude that vortices develop in the shear-banding flow of giant micelles. However, they do not allow the determination of the radial structure of the vortex flow. To address this question, we take advantage of the scattering contrast between both bands, due to concentration effects. Using white light illumination of the sample, free of seeding particles, we are able to directly image the cellular structures (Fig.~\ref{fig2}.g). The latter are detectable because of local index variations probably due to the concentration difference between the two phases. The counter-rotating vortices form concomitantly with the interface undulations and are stacked along the vorticity direction. They grow from the inflow regions and finally span the high shear rate band with a reduced extension in the low shear region when the steady-state is achieved (see Movie in Suplemental). \\
\begin{figure}[b]
\includegraphics{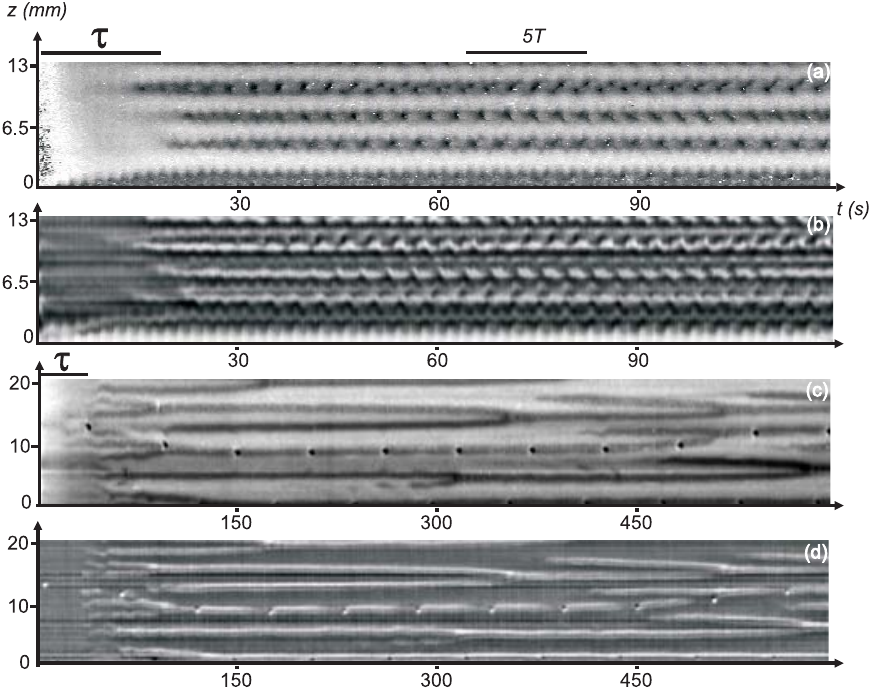}
\caption{Evolution in gray scale of the interface profile $I(z)$ and the reflected intensity distribution $I_o(z)$ as a function of time during step shear rate from rest to (a-b) $\dot\gamma$=35 s$^{-1}$ and (c-d) $\dot\gamma$=75 s$^{-1}$. The black and white dots observable in Figs.~3.c and d correspond to a bubble, trapped in a recirculation zone and that periodically crosses the observation fields.
\label{fig3}}
\end{figure}  
To investigate the role of the control parameter on the flow dynamics, we record the interface profile $I(z)$ and the reflected light distribution $I_o(z)$ as a function of time and construct in gray scale the corresponding spatiotemporal diagrams $I(z,t)$ and $I_o(z,t)$ . Note that the $I_o(z,t)$ diagrams (Fig.~\ref{fig3}.b and d) are fully reproduced by plotting as a function of time, an intensity profile along a vertical line in the flow cross section going through cellular structures in Figure 2.e. Fig.~\ref{fig3}.a and b show the patterns observed for  $\dot\gamma$=35 s$^{-1}$. After a transient of duration $\tau$ corresponding to the formation of a flat interface between bands, we observe the destabilization of the interface together with the development of axisymmetric cellular structures. Our experimental resolution do not allow to determine whether the Taylor-like velocity rolls are triggered by the interfacial instability or \textit{vice-versa}. The growth of the interface instability and the appearance of radial and axial components in the flow field are followed by saturation in a steady-state characterized by well-defined wavelength and amplitude of the interface, and a steady vortex structure. The patterns are only slightly modulated in time with a period $T$. This modulation corresponds to a small variation of amplitude in the inner crests (dark regions in Figs.~\ref{fig3}.a and c)) of the interface profile \cite{Lerouge2008} and produces a rope-like structure in the spatiotemporal sequence of the reflected light $I_{o}(z)$. It could result from periodic oscillations of the vortices boundaries. These extremely regular patterns are representative of the system dynamics on a large part of the stress plateau, for $\dot\gamma$ typically between 15$\pm$1 s$^{-1}$ and 60$\pm5$ s$^{-1}$. In this range of $\dot\gamma$, we find that the asymptotic wavelength of the patterns increases with the shear rate (Fig.~\ref{fig4}.a) while the period of the modulation keeps a constant value (Fig.~\ref{fig4}.b). For comparison, the wavelength of the spatial structure $I_{p}(z)$ obtained from PIV measurements has been added in Fig.~\ref{fig4}.a. For all shear rates, the height of a vortex pair and the interfacial wavelength are in good agreement. The circulation cells essentially span the high shear rate band in the radial direction, but their extension along the vorticity direction can reach about two gap widths as the control parameter is increased.\\
For higher values of the shear rate ($\dot\gamma >$ 65 s$^{-1}$), the system exhibits a more complex dynamics as illustrated in Fig.~\ref{fig3}.c and d. Once again, the dynamics of the tracers is remarkably correlated with the one of the interface. After the transient, the interface amplitude saturates but the wavelength continuously evolves with time due to nucleation and merging of inner crests. Since a pair of vortices develops between two consecutive inner crests, such a dynamics corresponds to a succession of creations and annihilations of pairs of vortices. 
\begin{figure}[t]
\includegraphics{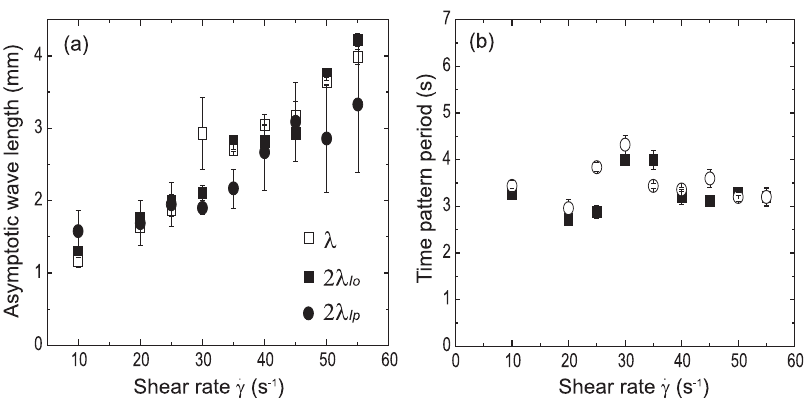}
\caption{(a) Comparison between the asymptotic  interfacial wavelength $\lambda$ ($\square$) and the height of a vortex pair inferred respectively from $I_o(z)$ ($\blacksquare$) and $I_p(z)$ ($\bullet$). (b) Period $T$ of the temporal modulation of the interface profile ($\circ$) and of the rope-like structure of the reflected intensity ($\blacksquare$). The error bars represent the standard deviation obtained over several experiments. The time and length scales are computed by direct Fourier transform. 
\label{fig4}}
\end{figure} 
These processes do not tend toward a stationary situation, rather leading to a chaotic dynamics. Note that at low shear rates (below 15 s$^{-1}$), the reduced size of the induced band prevents the description of the flow dynamics.  \\
The present results demonstrate that the shear-banding flow of our semi-dilute giant micelles solution is three-dimensional, far away from the classical picture usually used in this field \cite{Salmon2003,Hu2005}. Very recently, the formation of Taylor-like velocity rolls has been predicted by a nonlinear numerical study of the modified Johnson-Segalman model \cite{Fielding2007}~:~the 3D flow is triggered by the instability of the 1D gradient banded state with respect to interfacial undulations along the vorticity direction, with normal stress and shear rate jumps accross the interface as the underlying mechanism \cite{Fielding2005}. Some of our observations are qualitatively captured by this model. However, it does not reproduce the complex flow dynamics we report. Alternatively, another possible origin for the Taylor-like vortices is the elastic instability originally observed in dilute polymer solutions \cite{Larson90,Larson92} and more recently in rod-like systems \cite{Kang2006,LinGibson2004}. In working conditions where the streamlines are curved and the fluid is elastic, the base orthoradial flow can become unstable due to a centripetal body force called 'hoop stress' \cite{Larson90}. Such an hypothesis has been proposed to explain fast temporal fluctuations of the flow field that lead to suspicions for the presence of 3D shear-banded flow in a concentrated giant micelles system \cite{Becu2007}. Following the analysis in Ref.~\cite{Becu2007}, we compute a modified Weissenberg number $\Sigma=(e/R_i)^{1/2}Wi$ as a function of the applied shear rate $\dot\gamma$ . Here $Wi$ is an approximate Weissenberg number obtained from the stress ratio $N_1/\sigma$ \cite{Larson88}, and $(e/R_i)$ the gap ratio. The critical value $\Sigma_c$ above which an elastic instability appears is around 6 \cite{Larson90,Mckinley96}. This stability criterion has been established in the framework of the Oldroyd-B model. If we assume that each band can be modelled by an Oldroyd-B equation \cite{Kumar2000}, we can compare the modified Weissenberg number $\Sigma$ in each band with the critical value $\Sigma_c$. Note that, if the Oldroyd-B model does not describe shear-thinning properties, the shear-thinning character is supposed to reduce the instability threshold \cite{Larson94}. In the linear regime, $\Sigma=0$ and it is found to increase with $\dot\gamma$ in the nonlinear regime, namely above 1 s$^{-1}$ (data not shown). Along the first branch, $\Sigma$ never exceeds 0.5 while for $\dot\gamma\geq\dot\gamma_h$, $\Sigma>\Sigma_c$, suggesting that the induced band is likely to undergo an elastic instability, leading to formation of cellular structures.\\
In summary, our results provide the first direct evidence for the existence of Taylor-like velocity rolls in the shear-banding flow of wormlike micelles. Moreover, we showed that the dynamics of the Taylor-like  vortices is strongly correlated with the interface dynamics, leading to fundamental questions about the physical mechanism driving the three dimensional flow. Between the interface instability and the formation of Taylor-like vortices, one challenging task is now to determine which one of these phenomena drives the other, and under which physical mechanism. More generally, our results could potentially explain some fluctuating behaviors observed recently in giant micelles systems and open new perspectives towards the understanding of shear-induced instabilities and heterogeneous flows in complex fluids.
\section*{\large{Acknowledgements}}
The authors thanks A. and S. Asnacios, N. Biais, J.L. Counord, O. Dauchot and B. Ladoux for fruitful discussions, and the ANR JCJC-0020 for financial support.
\bibliographystyle{unsrt}

\end{document}